\documentstyle[12pt,fleqn]{article}
\renewcommand{\baselinestretch}{1.2}

\newcommand{\bea}{\begin{eqnarray}}
\newcommand{\beq}{\begin{equation}}
\newcommand{\eea}{\end{eqnarray}}
\newcommand{\eeq}{\end{equation}}
\newcommand{\nnu}{\nonumber}
\newcommand{\di}{\mbox{d}}
\newcommand{\spav}[1]{\parbox{1mm}{\vspace*{#1}}}

\begin{document}

\begin{titlepage}
\begin{flushright}
CERN-TH.6688/92
\end{flushright}
\begin{center}
{\Large\bf Testing Supersymmetry in Weak Decays}\\
{\Large\bf by Means of Time Reversal Invariance}\\
\spav{1.5cm}\\
{\large Ekaterina Christova}
\spav{1cm}\\
{\em Institute of
Nuclear Research and
 Nuclear Energy}\\
{\em Boul. Tzarigradsko
Chaussee 72, Sofia 1784, Bulgaria.}
\spav{1cm}\\and
\spav{1cm}\\
 {\large Marco Fabbrichesi}
\spav{1cm}\\
{\em CERN, Theory Division}\\
{\em CH-1211 Geneva 23, Switzerland}\\
\spav{1.5cm}\\
{\sc Abstract}
\end{center}
The minimal supersymmetric
 extension of the standard model allows for some of the
coupling strengths to be complex parameters. The presence
of such imaginary phases can lead to violations of time reversal
invariance, which can be tested if
correlations in products of
 an odd number of polarizations and momenta are measured
 and
found to be different from zero.
As an example, we consider the triple product
$\mbox{\bf J} \cdot \left(
\mbox{\bf p}_1 \times  \mbox{\bf p}_{2} \right)$  in the
$\beta$-decay of the neutron, of the $\Sigma^-$, and in
the decay $K^{+}_{3\mu}$. For these low-energy decays, we find
that the present
experimental precision is not enough to
 provide useful bounds on combinations of
such phases and the masses of the supersymmetric particles. At higher
energies, the same time reversal violating correlation in the
semileptonic decay of the $t$ quark is of the order of
$\alpha_s/\pi$, made bigger by the
large mass of the decaying quark.
\vfill
\spav{.5cm}\\
CERN-TH.6688/92\\
Revised version, May 1993

\end{titlepage}

\newpage
\setcounter{footnote}{0}
\setcounter{page}{1}

{\bf 1.} The presence of coupling strengths, which cannot be made real
 by a suitable redefinition
of the particle fields, is a feature of any, and in particular of the minimal,
supersymmetric extension of the standard model~\cite{MSSM}.

An immediate
 physical consequence of the irreducible complex nature of such a
Lagrangian
is that time reversal
invariance can be
violated---and CP conservation as well if, as we assume here, CPT
 invariance holds---to a larger degree
than  in the standard model.

Such a violation, if present, gives to otherwise vanishing observables
a finite size which
can be probed in low-energy experiments of sufficient precision. The first
example that comes to mind is, probably, a non-vanishing
electric dipole moment
 of the electron and the neutron~\cite{ed}.
A distinct possibility, and the one we consider in this letter,
is represented by
 correlations of  an odd number of polarizations and momenta, an example
of which is the triple product
\beq
\mbox{\bf J} \cdot \left( \mbox{\bf p}_1 \times
\mbox{\bf p}_2 \right) \, ,\label{1}
\eeq
measuring the polarization transverse to the plane of the two momenta.
The quantity~(\ref{1}) is odd under the action of the time reversal operator
$T$
because both the momenta $\mbox{\bf p}_i$
and the polarization {\bf J}
change sign; it is even under the parity operator $P$, which changes the sign
of
the momenta only.

Such $T$-odd correlations have been looked for in nuclear
physics~\cite{history} as clues to new dynamical laws and with substantial
improvements in the accuracy
of the measurements over the years. It is then
interesting, we believe,  to try to
establish how much of their size could in principle originate from the
minimal supersymmetric extension of the standard model.

This kind of enquiry, in which the supersymmetric
sector manifests itself only by radiative corrections in the
low-energy interactions of ordinary
particles,  trades
the high energies required in  actually producing the
supersymmetric partners for the high precision needed to probe their effect
in loop corrections.

On the other hand, because these $T$-odd correlations
induced by supersymmetry are  suppressed by powers of the
ratio between the mass of the
polarized particle and the
heaviest supersymmetric mass in the loop, it is interesting to look
also for semileptonic decays
at higher energies, where more massive particles are available. Here we have
in mind, above all, the
$t$ quark, the mass of which now seems to be of the same order as
the supersymmetric scale~\cite{CDF}. Therefore, the
production of these quarks at the LHC and SSC will
open new possibilities for
studying correlations such as~(\ref{1}) and
thus detect the effect, if any, of
supersymmetry.

\spav{1.5cm}\\
{\bf 2.} It is useful, before discussing any physical process, to
briefly recall to what
extent the non-vanishing of $T$-odd correlations such as~(\ref{1})
 is a genuine signal of
time reversal invariance violations. Because of the anti-unitary nature of
the time
reversal operator $T$~\cite{Wigner}, once loop amplitudes are included in the
computation, it is not necessarily true that $T$-odd observables violate
time reversal
invariance. The reason can be readily understood as follows. The $T$-odd
observable
originates, for example,
 in the interference between the imaginary part of a one-loop
amplitude and a tree diagram, which is real. However, such an imaginary
part can
occur in two distinct ways: either directly,
from the imaginary phases in the coupling
strengths, or, indirectly,  from the
imaginary part of the loop integral. This latter unitary
phase---sometime also referred to as the
final state interactions---contributes
to any $T$-odd observable
without representing a violation of time reversal
invariance and therefore constitutes
 a sort of
background that has to be taken into
account~\cite{FSI}.
As we shall see,
a direct estimate of these effects is possible.

We also notice that at the
supercolliders,  the semileptonic decays
we are interested in come together
with their $CP$ conjugate, as in the reaction
$q\bar{q} \rightarrow t\bar{t}$, which gives both
$bl^-\bar{\nu}_l$ and $\bar{b}l^+\nu_l$ as final states;
it is therefore possible to take the difference between
the two $CP$-conjugate cross sections in such a way that the unitary phase
background
(which is the same for the two decays) is eliminated~\cite{Golo}.

\spav{1.5cm}\\
{\bf 3.} Let us now consider the minimal supersymmetric
extension of the standard model.

We neglect generation mixing. Hence, only three
terms in the supersymmetric Lagrangian  can give
rise to $CP$-violating phases  which cannot be rotated
away~\cite{phases}: The superpotential
contains a complex coefficient $\mu$ in the term bilinear
in the Higgs superfields. The soft supersymmetry breaking operators
introduce two further complex terms, the gaugino masses
$\widetilde{M}_i$ and the left- and right-handed squark mixing
term $A_q$. We
consider only the latter two, which are carried by truly supersymmetric
particles, and leave out the additional contribution of the Higgs sector.

As long as we are not committed to any specific model of supersymmetry
breaking, it is not important to evaluate all possible diagrams  which
can give a contribution. Among them, we concentrate on the gluino Penguin
diagram, in which a gluino line is attached to two scalar-quark vertices
(see Fig.1). We leave out  contributions in
 which  neutralinos and charginos enter the loop, they
 are slightly suppressed by a factor $\alpha_w/\alpha_s$ with
respect to the gluino loop.

 The
 squark  mass eigenstates $\tilde{q}^j_{\alpha,n}$ are related to the weak
eigenstates $\tilde{q}^j_{\alpha,L}$ and $\tilde{q}^j_{\alpha,R}$
through the mixing matrix:
\bea
\tilde{q}^j_{\alpha,L} & = & \exp (-i\phi_{A_q}/2) \left[ \cos \theta\:
\tilde{q}^j_{\alpha,1} + \sin \theta \:   \tilde{q}^j_{\alpha,2} \right] =
\sum_m a^L_m \tilde{q}^j_{\alpha,m} \\
\tilde{q}^j_{\alpha,R} & = & \exp (i\phi_{A_q}/2) \left[ \cos \theta\:
\tilde{q}^j_{\alpha,2} - \sin \theta \:   \tilde{q}^j_{\alpha,1} \right] =
\sum_n a^R_n \tilde{q}^j_{\alpha,n}
\eea
where
\beq
\tan 2\theta = \frac{2|A_q|m_q}{(L^2 - R^2)\widetilde{m}} \, ,
\eeq
and
\beq
A_q \widetilde{m}
m_q = \xi_q v_2 + \mu^* h_q v_1 \qquad A_q = |A_q| \exp
i\phi_{A_q} \, . \eeq
$\widetilde{m}L$ and $\widetilde{m}R$ are the squark mass parameters,
$v_i$ the vacuum expectation values of the Higgses, and $\xi_q$ the
coefficient in the cubic term of the soft breaking operator. The
diagonalization of the squark masses gives the eigenvalues
\beq
\widetilde{m}_{1,2}^2 = \frac{1}{2} \left\{ (L^2 + R^2)\widetilde{m}^2
 + 2m_q^2 \mp
\left[ (L^2 - R^2)^2 \widetilde{m}^4 + 4m_q^2|A_q|^2
\widetilde{m}^2 \right]^{1/2}
\right\} \, . \eeq

 The gluino majorana mass
\beq
\widetilde{M}_g = \widetilde{m}_g \exp (i\phi_g)
\eeq
gives an additional phase shift once it has been rotated into the
interaction to make the masses  real.

The relevant terms in the Lagrangian are the following two:
\bea
L_{\tilde{q}\tilde{q}W} &=& -\frac{ig_w}{\sqrt{2}} W^-_\alpha \left(
\tilde{d}^*_L
\stackrel{\leftrightarrow}{\partial^\alpha} \tilde{u}_L \right) + h.c.
 \, ,  \\
L_{\tilde{q}q \tilde{g}} &=& \frac{g_s}{\sqrt{2}} T^a_{jk} \sum_{\alpha=u,d}
\left[  \bar{\tilde{g}}_a (1-\gamma_5) q_\alpha ^k
\Gamma_{L*}^m \tilde{q}^{j*}_{\alpha ,m} + \bar{q}_\alpha ^j (1+\gamma_5)
\tilde{g}_a \Gamma_{L}^m\tilde{q}^{k}_{\alpha ,m}  \right. \nnu \\ & & - \left.
\bar{\tilde{g}}_a (1+\gamma_5) q_\alpha ^k \Gamma_{R*}^n
\tilde{q}^{j*}_{\alpha
,n} - \bar{q}_\alpha ^j (1-\gamma_5) \tilde{g}_a
\Gamma_{R}^n\tilde{q}^{k}_{\alpha ,n} \right]  \label{c}
\eea
where
\beq
\Gamma_{L}^m = a_m^{L*}\exp ( -i\phi_g) \qquad
 \Gamma_{R}^n =a_n^{R}\exp (-i\phi_g) \, .
\eeq

 Both the $\Gamma_L^m$ and $\Gamma_R^n$ are
determined by the supersymmetry breaking mechanism and
 are in general complex
numbers~\cite{MSSM,phases} that cannot be made real by a redefinition of
the phases.
However, the presence of such an imaginary phase is not sufficient. Inasmuch as
we want to compute terms proportional to the polarization of one of the
external fermions, also
the chiral structure of the diagram is important.
It must be such that the fermion of which we measure the polarization changes
its helicity as compared to the tree level diagram.
The relevant diagrams are
depicted
in Fig.2, with  crosses representing the point of
chirality flip due to the mass term; there are then three of them.

Yet, these one-loop diagrams by themselves cannot contribute to a
correlation such as~(\ref{1})
because there are not enough independent momenta. In
fact, (\ref{1}) is nothing but the covariant quantity
\beq
\epsilon^{\alpha\beta\gamma\delta}J_{\alpha}p_{\beta}p_{\gamma}p_{\delta}
\label{eps}
\eeq
in the rest frame of one of the momenta. Clearly, (\ref{eps}) requires
at least three independent momenta.
It is therefore necessary to insert the diagrams
of Fig.1 in some decay or scattering
process.

To summarize: in order for a Feynman diagram to lead, by interference with the
tree diagram, to a time reversal
violation in the cross section,
 it must satisfy two
independent requirements: it must contain an imaginary phase and it must flip
chirality with respect to the tree level diagram.

\spav{1.5cm}\\
 {\bf 4.} We now turn to a specific example, namely the $\beta$-decay
of the neutron,
 which we would
like to analyze in some detail because it can be used as a template
for the other processes we are interested in.

Since we
are after new physics induced by the supersymmetric sector,
 we
must first of all
 estimate the potential contribution of the standard
model itself.
The imaginary phase, which comes from
the three-family  quark mixing in the corresponding
one-loop diagram, is suppressed by the unitary constraint, which makes
the leading contribution roughly  eight orders of
magnitude smaller than the current experimental bounds.

Next, it is important to isolate the unitary phase background.
It  originates in
the final state interactions, which in this case are only electromagnetic.
Even though one would
expect them to be of the order of $Z\alpha$, this turns out to
be incorrect because for
the standard model and its minimal
supersymmetric extension, and for any chiral theory,
 such a correction vanishes and the first unitary
phase comes only as a recoil effect  of order $Z\alpha E_e /M$,
where $M$ is the mass of the nucleon or of the nucleus. It has been estimated
to be $2.6 \times 10^{-4}$~\cite{FSI}.

Because the error
in the best experimental bound
 is still larger than the unitary phase background,
 we can compare the supersymmetric result
directly with the experiments.

Let us now
insert the diagrams of Fig.2 in the $\beta$-decay amplitude
\beq
n \rightarrow p e \bar{\nu}_{e} \, \label{2}.
\eeq
As we have pointed out, only the chiral
structure in which a right-handed  fermion ends up as a
left-handed one is needed.
The potentially relevant terms in the  vertex  are thus:
\bea
\bar{u}(p_p) \Gamma ^\mu u(p_n) & = &
\bar{u}(p_p)  \Bigl[ (1+\gamma_5)
 P^\mu {\cal A} + (1-\gamma_5)
 P^\mu {\cal B} \:
 \Bigr.  \nnu \\
& &  \Bigl. +\:
\gamma^\mu (1 + \gamma_5)\, {\cal C} \Bigr]  u(p_n)  , \label{G}
\eea
where $P\equiv p_p+p_n$.

Unfortunately, a direct evaluation of the hadronic matrix elements of
these operators is not possible. Naive dimensional
analysis~\cite{Georgi} suggests that they are of the same order as the
ones computed by means of quark matrix elements.

In the $\beta$-decay of the neutron all three operators are about of
the same order because of the closeness of the masses of the $u$ and
$d$ quark. The Penguin diagram gives
 \beq
{\cal A}  \equiv  -i\frac{g_w}{\sqrt{2}} g_s^2 \tilde{m}_g \sum_{m,n}
\Gamma_{L}^{m*}\Gamma_{L}^n\Gamma_{R}^{n*}\Gamma_{L}^m \left[ I_0^{m,n} -
I_1^{m,n} \right] \xi_A \label{A} \, ,
\eeq
\beq
{\cal B}  \equiv   -i\frac{g_w}{\sqrt{2}} g_s^2 \tilde{m}_g \sum_{m,n}
\Gamma_{L}^{n}\Gamma_{L}^{n*}\Gamma_{R}^{m}\Gamma_{L}^{m*} \left[
I_0^{m,n} - I_1^{m,n} \right] \xi_B \, ,  \label{B}
\eeq
and
\beq
{\cal C} \equiv  -i\frac{g_w}{\sqrt{2}} g_s^2  \sum_{m,n}
\Gamma_{R}^{n *}\Gamma_{L}^{m*}\Gamma_{R}^{m}\Gamma_{L}^{n} \:\:
I_2^{m,n} \xi_C \, . \label{C}
    \eeq

The coefficients $I_0^{m,n}$ and $I_1^{m,n}$, which appear
in~(\ref{A}) and (\ref{B}), depend on the
masses of the gluinos  and squarks (indices $n$ and $m$); they
 come from the integration over
the loop momentum and are defined as follows:
\bea
I_0^{m,n} & = & i \pi^2 \int_0^1 \di x x \int_0^1 \di y \frac{1}{M^2 - p^2}
\nnu \\
I_1^{m,n} & = & i \pi^2 \int_0^1 \di x x \int_0^1 \di y \frac{1-xy}{M^2 - p^2}
\, ,
\eea
where
\bea
p^\mu & = &  (1-x) p_u^\mu  + x(1-y) p_d^\mu \nnu \\
M^2 & = &
x(1-y) (m_{d}^2 - \widetilde{m}_n^2) - (m_{u}^2 - (1-x) \widetilde{m}_m^2)
 + xy \widetilde{m}_g^2  \, . \label{I}
\eea
Similarly, $I_2^{m,n}$ comes from the finite part of the vertex
renormalization and is given by
\beq
I^{m,n}_2 = i\pi^2 \int \di x x \int \di y \ln \frac{ M^2 - p^2}{4\pi
\mu^2} \, ,
\eeq
where $\mu$ is the renormalization point.

The coefficients $\xi$'s come from the renormalization group
running of these
operators in going from the supersymmetric scale---at which we
have computed the loop correction---to the nuclear scale. They can be
estimated by means of their
anomalous dimension~\cite{anomalous}. They are however
of order of one and we will simply carry a common factor $\xi$ to
indicate this correction.

 The relevant cross section can
now be computed and is \beq \frac{\di \Gamma}{\di E_e \di \cos \theta_{e
\nu_e}} = \frac{G_{\mu}^2}{\pi^3}
  E_e^2E_{\nu_e}^2
\left[ 1
 + D \, \mbox{\bf n} \cdot \frac{ \mbox{\bf p}_e \times
\mbox{\bf p}_{\nu_e}}{E_e E_{\nu_e}} \right]
  \label{crosssection} \, ,
\eeq
where $E_{\nu_e}= m_n - m_p -E_e$. In~(\ref{crosssection})
the first term is the leading
tree level contribution and the second one, the one
we are interested in, is
obtained by the interference of the one-loop amplitude and the tree. This term
violates time reversal invariance. The coefficient in front of it is defined to
be
 \beq
D \equiv \frac{2\sqrt{2}}{g_w} \Biggl\{ m_N  \mbox{Im}\,  {\cal A}
- m_N\mbox{Im}\,  {\cal B}  + \mbox{Im} {\cal C} \Biggl\}
\label{D} \eeq
with $\cal A$, $\cal B$ and $\cal C$ given
 respectively by~(\ref{A}),
(\ref{B}) and (\ref{C})
  in the
framework of the minimal supersymmetric
extension of the standard model.
 $\mbox{\bf n} \equiv$
{\bf J}/J points to the direction of the polarization of the
decaying neutron.

In order
to  obtain an overall estimate of the effect,
we
consider the case in which
the squark masses $\widetilde{m}_1$ and
 $\widetilde{m}_2$ are almost
degenerate and  equal to $\widetilde{m}_g \simeq \widetilde{M}$. Because of the
orthogonality of the coefficients $\Gamma^{L,R}$ we must however keep the
dependence on the masses before
summing over $m$ and $n$. This gives a factor
\beq
\frac{\widetilde{m}_2^2 -
\widetilde{m}_1^2}{\widetilde{M}^4} \simeq
\cos \theta \sin \theta \frac{2|A_q|m_q}{\widetilde{M}^3}
\eeq
in front of the integrals, that are now
dimensionless quantity that can  be computed. In the same approximation $\cos
\theta \sin \theta = 1/2$.

  This way, we obtain
\bea
D & \simeq & \frac{\alpha_s}{12\pi} \xi \left[ \left(
\frac{m_N m_d}{\widetilde{M}^2} \right)  |A_d|\sin
\left( \phi_{A_d}  -\phi_g \right)
 + \left(
\frac{m_N m_u}{\widetilde{M}^2} \right) |A_u|  \sin
\left( \phi_{A_u}
 -\phi_g \right)\right. \nnu \\
& &  + \left. \left(
\frac{m_u m_d}{\widetilde{M}^2} \right)  |A_u||A_d| \sin \left(
\phi_{A_d} - \phi_{A_u} \right) \right] \, . \label{CP} \eea

The result (\ref{CP})
can be compared with the best
experimental
bound available, which is~\cite{beta}:
\beq
D = (4 \pm 8) \times 10^{-4}\, ,
\eeq
obtained in the nuclear decay $^{19}$Ne $\rightarrow ^{19}$F.
For a typical supersymmetric mass of 100 GeV, and $\alpha_s \simeq
.1$, the estimate (\ref{CP}) is too small to be useful. Notice that
the phases $\phi_{A_d}  -\phi_g$ and $\phi_{A_u}  -\phi_g$ are
 already
constrained to be smaller than $10^{-3}$
 from bounds on the electric dipole moment of the neutron~\cite{bounds}.
The phase $\phi_{A_d} - \phi_{A_u}$ is unconstrained but the coefficient
in front of it is even more suppressed than the other two by the two
quark masses.

\spav{1.5cm}\\
{\bf 5.} As we have seen, the smallness of the $u$ and $d$ quark masses
makes the observable (\ref{1})  small. An improvement can be found in
the  weak decay of the strange baryons,
because of the larger mass of the $s$ quark. For example,  the
observable (\ref{1}) in  the decay
\beq
\Sigma^- \rightarrow n e^- \bar{\nu}_e
\eeq
has been studied~\cite{sigma}. The experimental bound is however far
from been as good as for the $\beta$-decay and reads:
\beq
D = .11 \pm .10 \, , \label{exp}
\eeq
to be compared to our estimate that in
this case is
\beq
D \simeq \frac{\alpha_s}{12\pi} \xi  \left(
\frac{m_\Sigma m_s}{\widetilde{M}^2} \right)  |A_s|\sin
\left( \phi_{A_s}  -\phi_g \right)
\eeq
which, even though it is almost one hundred times bigger than before,
it is still five order of magnitude  smaller than (\ref{exp}).

 \spav{1.5cm}\\
{\bf 6.} Another low-energy process
in which  supersymmetric loop corrections can
play a role is the  decay $K^{+}_{3\mu}$, in which
\beq
K \rightarrow \pi \mu \nu_{\mu} \, .
\eeq
The experimental bound on the coefficient of {\bf J}$_\mu \cdot \left(
\mbox{\bf p}_\mu \times \mbox{\bf p}_\pi \right)$ in this
case is~\cite{kappa}:
\beq
D = (-3.0 \pm 4.7) \times 10^{-3} \, .
\eeq
The supersymmetric effect we have computed is
\beq
\frac{\alpha_s}{12\pi} \xi \left(
\frac{m_K m_s}{\widetilde{M}^2} \right) |A_s|\sin
\left( \phi_{A_s}  -\phi_g \right) \, ,
\eeq
which is roughly three order of magnitude too small.
Nevertheless,
such a semileptonic decay will be
an interesting candidate for new bounds as there is going to be an
improvement in the experimental sensitiveness~\cite{improve}.

\spav{1.5cm}\\
{\bf 7.} Up to this point,
we have considered $T$-odd correlations of polarizations and momenta only
in low-energy processes to show that they are potentially interesting probes
of new physics beyond the standard model and, in particular, of its minimal
supersymmetric extension.

Such low-energy experiments should be pursued and
considered as complementary to accelerator physics because
a substantial improvement in the measurements
 could provide new and useful bounds on imaginary phases
and supersymmetric masses.

At the same time, our results show that
the decay of a heavier particle would have the advantage of
being less suppressed
by the mass ratio to which $D$ is
proportional.

The production of $t$ quarks at the LHC and SSC will
make it possible to study the semileptonic decay
\beq
t \rightarrow b\bar{l}\nu_l
\eeq
and therefore test time reversal invariance by means of the observable
\beq
\mbox{\bf J}_t \cdot \left( \mbox{\bf p}_l \times \mbox{\bf p}_{\nu_l}
 \right) \, , \label{31}
\eeq
as in the $\beta$-decay, where now {\bf J}$_t$ is the polarization of the $t$
quark~\cite{t}.

The unitary background for this process has been recently estimated to be
around $10^{-3} - 10^{-4}$ for $m_t =100-200$ GeV~\cite{Liu}.

The large mass of the $t$ quark greatly enhances
the effect of supersymmetry, whereas the standard model contribution
remains  negligible.
The computation of the supersymmetric contribution to~(\ref{31})
 proceeds---but for
replacing the $d$ by the $t$ quark
and the $u$ by the $b$ quark---along
the same lines as
for the $\beta$-decay but because of the small mass of the $b$ quark with
respect to the $t$, only the  operator proportional to
 $\cal A$ in  (\ref{G})
contributes. The $t$ quark is not
expected to hadronize before decaying.

  By retracing our steps in the previous sections,
we obtain that
\beq
D \simeq \frac{\alpha_s}{12\pi} \left( \frac{m_t}{\widetilde{M}}
\right)^2 \left[ 1 + \frac{E_\nu}{m_t} \left( 1 - \cos \vartheta \right)
\right] \left[ 1 - 2 \frac{2E_l}{m_t} \right]
 |A_t| \sin
\left( \phi_{A_t}
 -\phi_g \right)   \, ,  \label{CP2}
\eeq
where the terms in the square brackets come from the kinematics that cannot be
neglected as we did before for the low-energy processes, $\vartheta$ being the
angle between the lepton momenta.

The mass $m_t$ is larger than 94
GeV~\cite{CDF}
 and therefore~(\ref{CP2}) is  roughly $\alpha_s/\pi$ for maximal $CP$
violation; a result that makes
the decay of the $t$ quark a very promising candidate in the search for
supersymmetric physics beyond the standard model by means of time reversal
violations.

\spav{1.5cm}\\
E.C. would like to thank Alfred Bartl and
Walter Majorotto for helpful discussions and
 J. Ellis and the Theory
Group at CERN for  their kind
hospitality. Her work has been partially supported by the Bulgarian National
Science Foundation, Grant Ph-16. M.F. thanks Stefano Catani and Paolo Nason for
helpful discussions.

\newpage
\renewcommand{\baselinestretch}{1}

\newpage

\spav{4cm}

{\bf Fig. 1:} The gluino Penguin diagram.

{\bf Fig. 2:} The relevant  diagrams for, respectively, the term proportional
to ${\cal A}$, ${\cal C}$ and ${\cal B}$.
 Crosses denote the change of
chirality
because of the  mass operator

\end{document}